\documentclass[twocolumn,times]{aastex631}
\usepackage{mathrsfs}

\shorttitle{Frequency-resolved lags for continuum reverberation}
\shortauthors{Cackett et al.}

\begin{document}

\title{Frequency-resolved lags in UV/optical continuum reverberation mapping}

\correspondingauthor{Edward M. Cackett}
\email{ecackett@wayne.edu}

\author[0000-0002-8294-9281]{Edward M. Cackett}
\affiliation{Wayne State University, Department of Physics \& Astronomy, 666 W Hancock St, Detroit, MI 48201, USA}

\author[0000-0002-0572-9613]{Abderahmen Zoghbi}
\affiliation{Department of Astronomy, University of Michigan, 1085 South University Avenue, Ann Arbor, MI 48109, USA}

\author{Otho Ulrich}
\affiliation{Western Michigan University, Department of Physics, 1120 Everett Tower, Kalamazoo, MI 49008, USA}
\affiliation{Wayne State University, Department of Physics \& Astronomy, 666 W Hancock St, Detroit, MI 48201, USA}

\begin{abstract}
In recent years, continuum reverberation mapping involving high cadence UV/optical monitoring campaigns of nearby Active Galactic Nuclei has been used to infer the size of their accretion disks.  One of the main results from these campaigns has been that in many cases the accretion disks appear too large, by a factor of 2 -- 3, compared to standard models. Part of this may be due to diffuse continuum emission from the broad line region (BLR), which is indicated by excess lags around the Balmer jump. Standard cross correlation lag analysis techniques are usually used to just recover the peak or centroid lag and can not easily distinguish between reprocessing from the disk and BLR.  However, frequency-resolved lag analysis, where the lag is determined at each Fourier frequency, has the potential to separate out reprocessing on different size scales.  Here we present simulations to demonstrate the potential of this method and then apply a maximum likelihood approach to determine frequency-resolved lags in NGC 5548. We find that the lags in NGC 5548 generally decrease smoothly with increasing frequency, and are not easily described by accretion disk reprocessing alone.  The standard cross correlation lags are consistent with lags at frequencies lower than 0.1 day$^{-1}$, indicating they are dominated from reprocessing at size scales greater than $\sim$10 light days.  A combination of a more distant reprocessor, consistent with the BLR, along with a standard-sized accretion disk is more consistent with the observed lags than a larger disk alone.
\end{abstract}

\section{Introduction}

The accretion disk \citep{shakurasunyaev}, X-ray corona \citep{haardt91}, and other regions surrounding the central supermassive black hole in Active Galactic Nuclei (AGNs) are beyond current spatial resolution limits for the vast majority of objects. To overcome this, the reverberation mapping technique uses time lags between light curves observed at different wavelengths to infer the geometry and kinematics of these inner regions \citep[see][for a recent review]{cackett21}.  The time lags are assumed to be due to light travel time, with the path length to the observer different for different regions.  While reverberation mapping was first used to determine the size scale of the H$\beta$-emitting broad line region (BLR), more recently there has been significant effort to measure the size and temperature profile of accretion disks.

If a central ionizing source (often assumed to be the X-ray corona) irradiates the accretion disk, then variations in the irradiating source can drive variations in the heating of the disk where it is thermally reprocessed and reemitted. The hotter, inner part of the disk will see the variations before the cooler outer part of the disk, leading to correlated continuum light curves with the UV leading the optical.  A standard \citet{shakurasunyaev} accretion disk with temperature profile $T \propto R^{-3/4}$ will give rise to wavelength-dependent lags, $\tau$, following $\tau \propto \lambda^{4/3}$.  See \citet{cackett07}, and references therein, for a more detailed discussion.

UV/optical continuum light curves have long been known to be well-correlated, and to have short interband lags of a few days, at most \citep[e.g.,][]{stirpe94,wanders97,collier98,sergeev05}.  Recent high cadence (better than once per day) observations combining Swift and ground-based observations have significantly improved the continuum lag measurements \citep{edelson15,edelson19,fausnaugh16,mchardy18,cackett18,cackett20,hernandezsantisteban20,kara21}. These campaigns have found common results: that the lags approximately follow $\tau \propto \lambda^{4/3}$; that the normalization of this relation is larger than expected for a standard disk; that lags in the $u$ and $U$ bands are generally systematically longer than expected based on an extrapolation of the other UV/optical lags; and that there is not a consistent relation between the X-rays and the UV/optical.

That the lags in the $u/U$ band are offset by approximately a factor of 2 \citep{edelson19}, was further highlighted by HST spectroscopic monitoring of NGC 4593 which resolved the continuum lags in this region of the spectrum, showing a clear discontinuity at Balmer jump \citep{cackett18}.  These results can be explained if there is significant additional continuum emission not from the disk but from the BLR \citep{koristagoad01,koristagoad19,lawther18,netzer20,netzer21}.  This diffuse continuum from the BLR should emit across the full UV/optical bands, and can affect the lags in all wavelengths.  The lag from this component increases with wavelength, except for discontinuities at the Balmer and Paschen jumps, and it is therefore hard to cleanly separate out its contribution \citep{koristagoad01}.

The spatial separation of the accretion disk and BLR continuum emitting regions should lead to the signals being distinguishable on different timescales, with disk reverberation taking place on timescales of a few days and BLR continuum reverberation taking place on timescales of weeks or longer.  In principle, filtering the light curves on these timescales can be used to pick out the different signals.  For instance, techniques such as subtracting off a moving box-car average have been used in some cases to remove long-term variability, essentially acting as a high-pass filter \citep[e.g.,][]{mchardy14,mchardy18,pahari20,vincentelli21}.  Other approaches are to implement more sophisticated time lag analysis.  For instance, one can use maximum entropy techniques to try and recover the response function through fitting the light curves \citep{horne94}. This has been used to show that in NGC 4593 the response function has both a prompt response and a tail to longer lags, that could be interpreted as signals from both the disk and the BLR \citep{mchardy18}.  Alternatively, \citet{chelouche19} use a bivariate reverberation model to suggest that the lags in Mrk~279 are dominated by the BLR. 

Frequency-resolved lag analysis is yet another approach.  This is widely used to perform X-ray reverberation \citep[see][for a review]{uttley14}, where it has been able to successfully separate out lags from different processes taking place on different timescales \citep[e.g.,][]{kara13}.  However, in X-ray analysis typically the observation lengths are significantly longer than the timescale of interest, providing a continuous time series that can be relatively simply analyzed with a fast Fourier transform to get lags at different frequencies.  The case in UV/optical reverberation is, of course, different with time series built up over many months but sampled irregularly, and often with gaps due to poor weather at the ground-based observatories.   However, several techniques have been developed in order to deal with gaps in light curves and still perform a frequency-resolved analysis.  Not all X-ray satellites provide long continuous exposures, thus \citet{zoghbi13} followed the method described by \citet{miller10a} to develop a maximum likelihood approach to determining power spectra and lags from unevenly sampled light curves.  Other approaches, such as using Gaussian processes \citep{wilkins19}, can similarly calculate frequency-resolved lags.

Here, we begin by presenting simulations to demonstrate the ability of frequency-resolved analysis to identify reprocessing from an extended region, before we apply the technique of \citet{zoghbi13} to analyze the light curves from the AGN STORM campaign on NGC 5548 and calculate frequency-resolved lags and discuss their implications.

\section{Frequency-resolved lags}

In a linearized reverberation mapping model, we can relate the reverberating (reprocessed) light curve, $r(t)$, to the driving light curve, $d(t)$, via
\begin{equation}
r(t) = \int_{-\infty}^{\infty} \psi(\tau) d(t - \tau) \; d\tau
\end{equation}
where $\psi(\tau)$ is the impulse response function (often called the transfer function).  This response function encompasses the information about the geometry of the reprocessor, and gives the response of the gas to a delta-function flare in the driving continuum.  The reprocessed light curve is a blurred and shifted (lagged) version of the driving light curve.  Typical time domain analysis performed in optical AGN studies calculate the cross correlation function (CCF), which is a convolution of the response function with the auto correlation function (ACF) of the driving light curve.  Thus measuring the centroid of the CCF approximately measures the centroid of the response function.  If the reprocessing region is complex, this will be reflected in the shape of the CCF.  For instance, a skewed CCF whose peak and centroid do not align might be indicative of an extended response function.  But, most CCF analysis does not attempt to study these effects and stops at measuring the lag via either the peak and/or centroid.

In the Fourier domain (indicated below using upper case symbols), the Fourier transform of $r(t)$ is simply the Fourier transform of $d(t)$ multiplied by the Fourier transform of the response, i.e.,
\begin{equation}
R(\nu) = \Psi(\nu) D(\nu) \;
\end{equation}
where $\nu$ is the Fourier frequency.  We can therefore also relate the power spectral density (PSD) of the reprocessed light curve to the driving light curve (see section 2.4 in \citealt{uttley14} and also \citealt{papadakis16}), with the reprocessed PSD equal to the PSD of the driving light curve multiplied by the modulus-squared of the Fourier transform of the response function:
\begin{equation}
|R(\nu)|^2 = \Psi^*(\nu) D^*(\nu) \Psi(\nu) D(\nu) = |\Psi(\nu)|^2 |D(\nu)|^2
\end{equation}  
where $^*$ indicates a complex conjugate.  To calculate the lag between two light curves in the Fourier domain we use the cross-spectrum, $C(\nu)$, which is the complex conjugate of the Fourier transform of one light curve multiplied by the Fourier transform of the other.  Thus, 
\begin{equation}
C(\nu) = D^*(\nu) R(\nu) = D^*(\nu) D(\nu) \Psi(\nu) = |(D(\nu)|^2 \Psi(\nu) 
\end{equation}
In other words, the cross-spectrum is equal to the driving signal PSD multiplied by the Fourier transform of the response function. The phase of the cross-spectrum gives the phase lag between the light curves at each Fourier frequency \citep[see sections 2.1.2 and 2.4.3 of][for a more detailed description]{uttley14}.  This phase lag can be converted to a time lag by dividing by $2\pi\nu$.  The cross-spectrum can therefore be used to determine lag as a function of frequency, providing more detail about the response function than simply measuring the average value, and allowing an investigation of the timescale (frequency) dependence of the response.

Reverberation acts to smooth light curves and, as described above, therefore changes the light curve's PSD.   The normalization of the PSD should drop, and its slope should steepen with the amplitude reduced most at the highest frequencies.  Qualitatively this can be easily seen looking at the light curves, with longer wavelength light curves (that are thought to originate from further out in the accretion disk) being smoother and having lower variability amplitudes \citep[for one example see fig.~2 in][]{fausnaugh16}.  A quantitative analysis by \citet{panagiotou20} measures the PSD of these NGC 5548 light curves, finding that they are all well described by a simple power-law.  The slope of the UV/optical PSDs are the same, but steeper than the X-ray PSD.  The normalization of the PSD decreases with wavelength, consistent with thermal reverberation.  However, there has not been a study of the cross-spectrum and frequency-resolved lags in AGN light curves across the full UV/optical range. The only previous attempt compared the X-ray and UV light curves in Mrk~335, and only saw tentative lags on long timescales, with all values consistent with zero within 2$\sigma$ \citep{griffiths21}.

\section{Simulations}\label{sec:sims}

\begin{figure*}
\centering
\includegraphics[width=0.8\textwidth]{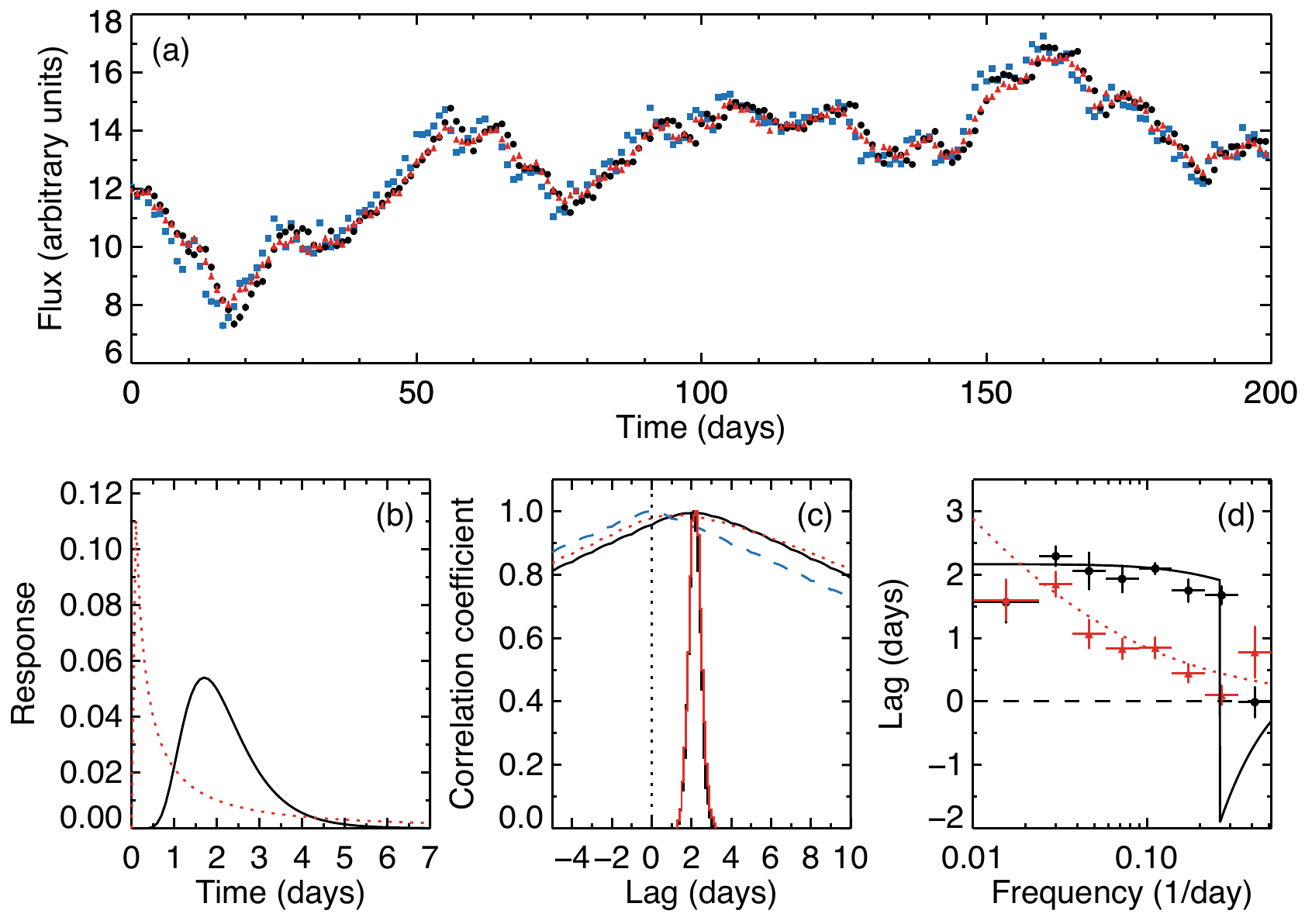}
\caption{(a) Simulated driving light curve (blue squares), and reprocessed light curves for response 1 (black circles) and response 2 (red triangles). (b) Response function 1 (black solid line) and response function 2 (red dotted line). Response 2 peaks at shorter times, but has a more prominent tail at late times. (c) Auto correlation function of the driving light curve (blue dashed line) compared to the cross correlation functions for response 1 (black line) and response 2 (red dotted line).  The black and red histograms show the centroid lag distributions from the FR/RSS procedure used to determine the lag uncertainties.  The centroid lags for response 1 and 2 are consistent within 1$\sigma$. (d) The frequency-resolved lags recovered from the light curves for response 1 (black circles) and response 2 (red triangles), with the model frequency-resolved lags shown as black solid and red dotted lines for response 1 and 2 respectively.}
\label{fig:sims}
\end{figure*}

To demonstrate the potential advantages of undertaking a frequency-resolved lag analysis, we perform some simple simulations.  We generate a driving light curve several times longer than needed \citep[to account for red noise leakage;][]{uttley02}, using the \citet{timmerkonig} algorithm and assuming a PSD with slope $-2$.  We then convolve this with a response function.  After this we cut the light curve down to 200 days, and sample with a cadence of once per day, adding random Gaussian noise of 1\% (panel (a) of Fig.~\ref{fig:sims} shows the light curves used).  We test a couple of different response functions, chosen so that their recovered centroid lag values are approximately the same, however their different properties mean that one has a more prominent tail at long delays.  Both responses are chosen to be log-normal functions:
\begin{equation}
 \psi (t) = \frac{1}{S\sqrt{2\pi} t} \exp^{-(\ln t - M)^2/(2S^2)}
 \label{eq:lgnorm}
\end{equation}
where $t$ is time in days.  For response 1 we choose $M = \ln{2}$ and $S = 0.4$.  For the second response (response 2) we also use a log-normal distribution, this time with $M = 0$ and $S = 2.0$.  The smaller $M$ means that the response peaks at shorter lags, but the larger $S$ broadens the distribution, giving a significant tail to long lags (as can be seen in panel (b) of Fig.~\ref{fig:sims}).  The peak and centroid of the first response are quite similar with 1.70 and 2.17 days respectively, the second response however, has a peak of 0.02 days and a centroid of 6.85 days.  The responses are calculated over the range $t = 0$~--~$1000$ days.

With these simulated light curves we test how well-distinguished the two response functions are by the standard CCF and frequency-resolved lags approaches.  First, we calculate the CCF with respect to the driving light curve and use the standard flux randomization, random subset sampling method \citep[FR/RSS;][]{white94,petersonetal04} to determine the peak and centroid lags and their uncertainties.  Panel (c) of Fig.~\ref{fig:sims} shows the CCFs and histograms of centroid values from the FR/RSS technique. By eye the CCFs look quite similar, though it can be seen that response 2 peaks at slightly smaller lags and has a higher correlation coefficient at larger lags compared to response 1.  Quantitatively, the recovered centroid lags are consistent within their 1$\sigma$ uncertainties, but the peak lags differ.  The recovered peak and centroid lags are $\tau_{\rm peak, 1} = 2.20 \pm 0.10$~days, $\tau_{\rm cent, 1} = 2.16 \pm 0.29$~days for response 1 and $\tau_{\rm peak, 2} = 1.35 \pm 0.10$~days, $\tau_{\rm cent, 2} = 2.19 \pm 0.31$~days for response 2.  The peak and centroid for response 2 are significantly different from each other, indicating the response function is asymmetric with a tail to long lags.  However, in most reverberation analyses the peak and centroid lags are usually simply quoted and rarely are the differences in the peak and centroid lags or the shape of the CCF investigated in more detail.

The differences between the two responses becomes clearer when looking at the frequency-resolved lags.  Since the simulations use a strict 1-day cadence, the Fourier analysis can be performed using standard fast Fourier transform techniques \citep{uttley14} to calculate the cross spectrum (which is then binned in frequency), and from the phase of the (binned) cross spectrum calculate the time lag.  The resultant lags as a function of Fourier frequency are shown in panel (d) of Fig.~\ref{fig:sims}, where we also show the predicted lags based on the model response functions.  Here, a clear difference between the frequency-resolved lags is easily seen.  

There are notable features in the lag-frequency plot worth describing.  Phase-wrapping (the sudden flip in sign of the lag) occurs as the phase lags are limited to the range $-\pi$ to $\pi$.  So, a positive or negative shift of half a wave cannot be distinguished.  For a symmetric response function this will occur at a frequency of $\nu = 1/2\tau_0$ (see section 4.1.1 of \citealt{uttley14}), where $\tau_0$ is the centroid of the response function.  For an asymmetric response function $\tau_0$ is skewed to slightly longer than the peak, and, of course, depends on the Fourier transform of the response function.  This phase-wrapping therefore indicates the approximate lag where the response function peaks.   Phase-wrapping is clearly seen in response 1, at a frequency of approximately 0.25 day$^{-1}$, corresponding  $\tau_0 = 2$ days.  For the second response the phase-wrapping occurs at a frequency higher than probed by the cadence of the data sampling.

Aside from the phase-wrapping the evolution in lag with frequency is also significantly different for the different models.  For response 1 at frequencies lower than where phase-wrapping occurs the lag remains approximately constant.  This flattening indicates that there is not significant additional response on long timescales.  For response 2, however, the short peak and long tail to the response causes the lags to smoothly decrease with increasing frequency -- in other words as you go to greater distances (longer lags) there is additional reprocessing.

At the lowest frequencies once you get beyond the equivalent timescale where the response function has dropped to zero, the lag becomes a constant and corresponds to the centroid lag of the response function.  This is clearly seen in panel (d) of Fig.~\ref{fig:sims} for response 1 (black line).  However, for response 2 (red dotted line) the 200-day campaign is not long enough to get frequencies low enough to completely reach the flat part of the lag-frequency evolution.  The length of the campaign obviously limits the lowest frequency bins that can be analyzed, and biases can be introduced if there are not enough Fourier frequencies per bin \citep[see e.g.,][]{epitropakis16}.  If there is response from a distant reprocessor, a 200-day campaign is not necessarily long enough to fully recover the lags. This can only be mitigated by having significantly longer monitoring campaigns, which presents a challenge given ground-based visibility of objects.

In summary, while with standard CCF approaches the recovered centroid lags are consistent between the two different responses used in the simulations, when looking at frequency-resolved lags the evolution of the lag with frequency allows to clearly distinguish between the two responses.  This demonstrates the potential in using the technique to better constrain the shape of the response function in UV/optical continuum reverberation and help determine the contribution to the lags from the accretion disk, broad-line region or other location.

\section{Frequency-resolved lags in NGC 5548}

The simulations demonstrate the potential power in frequency-resolved analysis to distinguish between different shaped response functions.  We therefore apply the frequency-resolved lags method to a real dataset.  Over the last 5 years or so a number of intensive, multi-band continuum reverberation mapping campaigns have taken place using Swift to anchor the analysis with high cadence (usually multiple times per day) X-ray and UV/optical light curves often spanning several hundred days.  AGN STORM \citep{derosa15} was the first campaign to combine Swift, Hubble Space Telescope and ground-based monitoring and clearly detects wavelength-dependent lags \citep{edelson15,fausnaugh16}.  With 262 epochs over 169 days in the Swift UVW2 light curve and 171 epochs over 175 days in the HST light curves it remains one of the best datasets currently available.  We therefore use these data to test this analysis method.  Since these real data are not evenly sampled in time we cannot use standard fast Fourier transforms, and so instead use a maximum likelihood method.

The lags are calculated using the maximum likelihood method outlined in \citet{zoghbi13}, which was first presented in \citet{miller10a}. The codes used for the analysis can be found at \url{https://zenodo.org/record/5566974} \citep{zoghbi21}.  A likelihood function for the observed light curves is constructed assuming that the power spectra and the response function are piecewise functions of Fourier frequency, and whose parameters are estimated by numerically maximizing the likelihood. A Gaussian likelihood is used, with a covariance matrix that is built from the Fourier transform of the piecewise functions.

In other words, the autocovariance function is defined as:
\begin{equation}
\mathcal{A} (\tau) = {\displaystyle \int |D(\nu)|^2 {\rm cos}(2\pi \nu \tau)\,d\nu}
\end{equation}
where $\nu$ is the Fourier frequency, $\tau$ is the time difference between pairs of time points, and $|D(\nu)|^2$ is the power spectral density, modeled here as a piecewise sum over $n_{\nu}$ frequency bins: $|D(\nu)|^2 = \sum_i D_i$. Likewise, the cross-covariance is defined as:

\begin{equation}
\mathcal{X} (\tau) = {\displaystyle \int |D(\nu)|^2 |\Psi(\nu)| {\rm cos}(2\pi \nu \tau - \phi(\nu) )\,d\nu}
\end{equation}
where $|\Psi(\nu)|$ and $\phi(\nu)$ are the amplitude and phase of the response function. These two are also modeled as piecewise functions of frequency ($\Psi(\nu) = \sum_i \Psi_i$ and $\phi(\nu) = \sum_i \phi_i$). Assuming the observed light curves are generated by such general Gaussian Processes, a likelihood function can be constructed, and maximized to obtain best values $\Psi_i$ and $\phi_i$ as a function of frequency. The time lag is then obtained by dividing the phase $\phi_i$ by $2\pi \nu_i$. All the lags are measured relative to the light curve at 1158\AA, which we take as the reference.

We report lags in 6 logarithmically-spaced frequency bins between 0.012 and 1.11 ${\rm days}^{-1}$ in Table~\ref{tab:lags}. Two additional frequency bins at the start and end of this range ($0.0017-0.012$ and $1.11-2700$ ${\rm days}^{-1}$) were included in the maximum likelihood calculations, but were not in the  modeling because not all light curves have information in this range, and also because the first and last bin tend to show biases in their measured lags (\citealt{zoghbi21}, see also \citealt{epitropakis16}). 

\begin{deluxetable*}{lCCCCC}
\tabletypesize{\small}
\tablewidth{0pt}
\tablecolumns{6}
\tablecaption{Lags (measured with respect to 1158\AA) in each Fourier frequency bin \label{tab:lags}}
\tablehead{ \colhead{Filter/} & \multicolumn{5}{c}{Lag (days) in Fourier frequency range}\\
\colhead{Wavelength} & \colhead{0.012 -- 0.025 day$^{-1}$} & \colhead{0.025 -- 0.054 day$^{-1}$}& \colhead{0.054 -- 0.115 day$^{-1}$}& \colhead{0.115 -- 0.244 day$^{-1}$}& \colhead{0.244 -- 1.111 day$^{-1}$}
}
\startdata
1367\AA & 0.59\pm0.19 & 0.20\pm0.10 & 0.06\pm0.10 & 0.10\pm0.08 & -0.03\pm0.38 \\
1479\AA & 0.17\pm0.27 & 0.42\pm0.11 & 0.17\pm0.11 & 0.09\pm0.09 & 0.04\pm0.14 \\
1746\AA & 0.77\pm0.28 & 0.37\pm0.14 & 0.33\pm0.13 & 0.25\pm0.12 & 0.00\pm0.31 \\
UVW2    & 1.41\pm0.39 & 0.10\pm0.20 & -0.12\pm0.18 & -0.23\pm0.14 & 0.01\pm0.48 \\
UVM2    & 1.47\pm0.59 & -0.02\pm0.22 & -0.03\pm0.23 & -0.18\pm0.16 & 0.05\pm0.46 \\
UVW1    & 1.41\pm0.58 & 0.32\pm0.26 & 0.11\pm0.22 & -0.13\pm0.20 & 0.05\pm0.45 \\
U$_{\rm Swift}$ & 2.54\pm0.68 & 1.08 \pm 0.31 & 0.36\pm0.39 & 0.24 \pm0.44 & -0.05\pm0.50\\
u             & 4.12\pm0.51 & 1.66\pm0.38 & 1.65\pm0.49 & 0.57\pm0.46 & 0.12\pm0.25\\
B            & 3.85\pm0.46 & 1.79\pm0.21& 1.09\pm0.42 & \multicolumn{2}{c}{$0.19\pm0.16$} \\
B$_{\rm Swift}$ & 1.94\pm0.88 & 0.89\pm0.44 & 0.72\pm1.69 &  \multicolumn{2}{c}{$0.10\pm0.66$} \\
g            & 3.48\pm0.57 & 1.73\pm0.34 & 1.61\pm0.31 & 0.70\pm0.37 & 0.02\pm0.27\\
V           & 3.74\pm0.39 & 2.72\pm0.20 & 0.95\pm0.22 & 0.65\pm0.28 & 0.04\pm0.47 \\
r            & 5.74\pm0.51 & 2.75\pm0.38 & 2.62\pm1.29 & -0.18\pm1.45 & 0.22\pm0.41 \\
R           & 4.25\pm0.71 & 3.13\pm0.37 & 1.55\pm1.01 & 1.33\pm0.84 & -0.02\pm0.48 \\
i             & 5.68\pm0.55 & 3.64\pm0.33 & 2.55\pm0.40 & \multicolumn{2}{c}{$0.36\pm0.39$} \\
I             & 5.57\pm0.67 & 3.15\pm0.41 & 0.42\pm3.07 & \multicolumn{2}{c}{$0.31\pm0.26$}\\
z            & 4.96\pm0.65 & 4.07\pm0.35 & 3.09\pm0.50 & -0.10\pm1.49 & -0.03\pm0.17
\enddata
\end{deluxetable*}

\begin{figure*}
\centering
\includegraphics[width=0.95\textwidth]{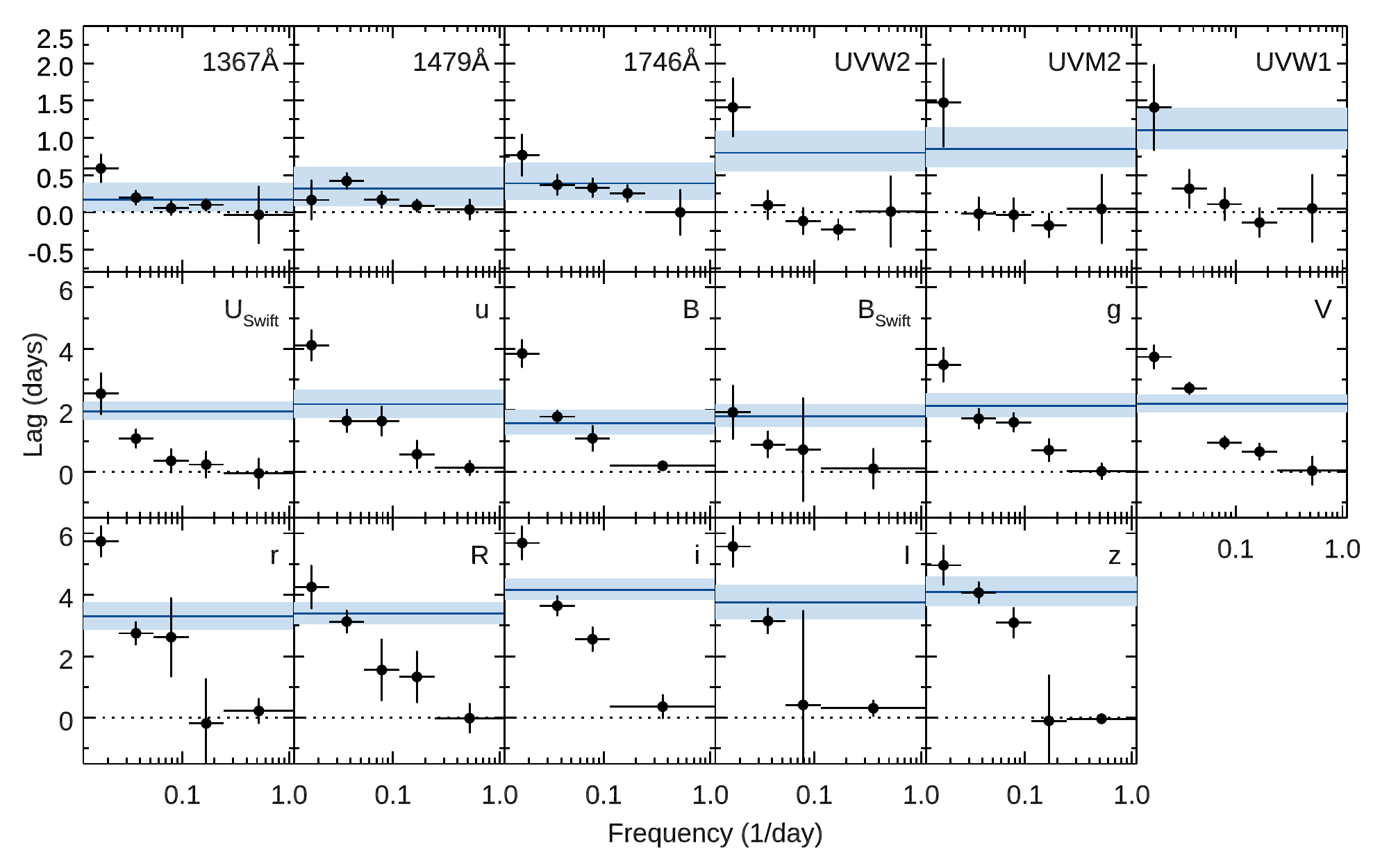}
\caption{Frequency-resolved lags in NGC 5548.  Blue lines are the CCF lags from \citet{fausnaugh16} for comparison. Note that the y-axis range in the top panel is narrower than for the other two panels to better visualize the shorter lags in those bands.}
\label{fig:lagfreq}
\end{figure*}

The uncertainties in the reported parameters are obtained by running Monte Carlo Markov Chains starting with a random set of parameters around the best fit obtained by numerical maximization of the likelihood function. We use the $1\sigma$ uncertainty of the lag as the standard deviation of the parameter chain. To allow subsequent modeling, we ensured that all measured lags have Gaussian probability densities. If any of the frequency bins is not consistent with a Gaussian distribution, we group it with a neighboring frequency to increase the signal and ensure normal uncertainties. As a consequence, we combined the two highest frequency bins in all light curves (resulting in 5 frequency bins), and for the lower cadence and/or signal light curves we end up with 4 frequency bins.  The recovered frequency-resolved lags are shown in Fig.~\ref{fig:lagfreq}. 

\begin{figure*}
\centering
\includegraphics[width=\textwidth]{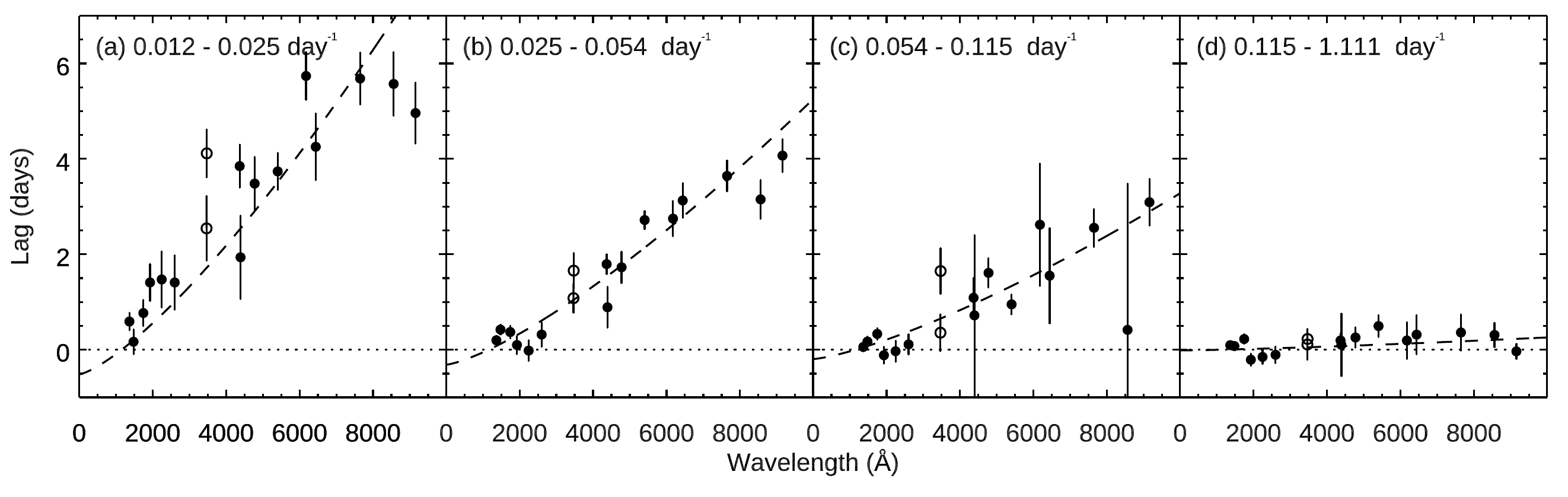}
\caption{Lags as a function of wavelength for each of the frequency bins.  The dashed line is the best-fitting $\tau \propto \lambda^{4/3}$ relation. The Swift $U$ and ground-based $u$ bands are shown as open circles.} 
\label{fig:laglambda}
\end{figure*}

For comparison we also show the cross-correlation lags of \citet{fausnaugh16} in Fig.~\ref{fig:lagfreq} as a blue horizontal region.  The cross-correlation lags are most consistent with frequency-resolved lags at frequencies lower than 0.1 day$^{-1}$, with the frequency bin from 0.025 -- 0.054 day$^{-1}$ often consistent with the cross-correlation lag.  In Fig.~\ref{fig:laglambda} we show the lags as a function of wavelength in each of the frequency bins. For wavebands with 5 frequency bins we average the lags between the two highest frequency bins in order to be able to compare with the wavebands with 4 frequency bins.   The lags in the three lowest frequency bins all approximately follow $\tau \propto \lambda^{4/3}$ (dashed line), while the highest frequency bin is dramatically different, with most lags consistent with zero.

The best-fitting $\tau = \tau_0\left(\lambda/\lambda_0 - 1\right)^{4/3}$ from the cross correlation analysis of \citet{fausnaugh16} gives $\tau_0 = 0.42\pm0.02$ days for their reference band of 1367\AA.  Adjusting to 1158\AA\ by subtracting the 1158\AA\ to 1367\AA\ lag of $0.17\pm0.16$ days gives $\tau_0 = 0.25\pm0.16$ days, consistent with the $\tau_0 = 0.32\pm0.01$ days we get from fitting the lags in the 0.025 -- 0.054 day$^{-1}$ frequency bin.  This further demonstrates how the CCF analysis is picking up the lags at low frequencies, corresponding approximately to size scales greater than 18 light days.  Given the red noise nature of AGN variability the light curves have the most power at the lowest frequencies. Filtering out the long time scale variability in the light curves (i.e. by subtracting a running box-car average) leads to shorter CCF lags.

\section{Modeling}

A clear trend is seen when looking at the lag-frequency plots for each waveband shown in Fig.~\ref{fig:lagfreq}.  Generally, they show a smoothly decreasing lag with increasing frequency.  The lag is consistently  longest at the lowest frequency examined, and close to zero above a frequency of 0.1 day$^{-1}$ in most bands.  Comparing to the simulations presented in Section~\ref{sec:sims} we can immediately rule-out a close-to symmetric response function.  Such a response function should show an almost flat lag-frequency spectrum up to the frequency where phase-wrapping occurs.  For instance, a centroid lag of 2 days (e.g., similar to the g-band lag) should show phase wrapping at 0.25 day$^{-1}$.  Since 0.25 day$^{-1}$  is a higher frequency than probed here we would expect a lag of 2 days in all frequency bins.  The simulations also show that a smooth decrease in lag with increasing frequency results from an asymmetric response function with a long tail at larger lags.  This immediately suggests that the continuum lags we observe come from an extended reprocessing region.  The fact that the lags are very short at frequencies greater than 0.1 day$^{-1}$, suggests reprocessing on size scales less than 10 light days happens very rapidly.

We note that accretion disk response functions are typically not symmetric and do show a tail at long lags.  For instance, see examples in \citet{cackett07}, \citet{starkey16} and \citet{kammoun19}.  Thus, to test whether accretion disk reprocessing models show the observed lag-frequency spectrum we use the models described in detail in \citet{cackett07}.  Those models are characterized by the disk temperature in the bright and faint state, $T_B$ and $T_F$.  We set $T_B$ and $T_F$ so that the centroid lag approximately follows the observed CCF lag centroids from \citet{fausnaugh16}, getting a good match with $T_B = 20900$~K and $T_F = 16800$~K.  We assume an inclination of $45^\circ$.

To calculate the model lags we note that we have calculated the observed lags with respect to the 1158\AA\ band.  In the disk reprocessing model this is not the driving light curve itself, and is itself a reprocessed light curve.  To correct for this we note that if the light curve in band A is $a(t)$ and has response function $\psi_A(t)$, and the light curve in band B is $b(t)$ and has response $\psi_B(t)$ then the Fourier transform of each light curve is:
\begin{eqnarray}
A(\nu) = \Psi_A(\nu)D(\nu) \\
B(\nu) = \Psi_B(\nu)D(\nu)
\end{eqnarray}
The cross spectrum of light curve B with respect to A will then be:
\begin{eqnarray}
C(\nu) = A^*(\nu) B(\nu) = |D(\nu)|^2 \Psi_A^*(\nu) \Psi_B(\nu) \; .
\end{eqnarray}
Since $|D(\nu)|^2$ is a set of real constants (and therefore does not affect the complex phase), the lags can be calculated from the phase of $\Psi_A^*(\nu) \Psi_B(\nu)$ alone.  Here, we adopt the 1158\AA\ response as $\psi_A(t)$, and the response for each band in turn is $\psi_B(t)$.  Since the lag between the driving light curve and 1158\AA\ is very short in these models (approximately 0.2 days), this is only a minor adjustment to the lags. It would be significantly more important if an optical band was used as the reference.

\begin{figure*}
\centering
\includegraphics[width=0.99\textwidth]{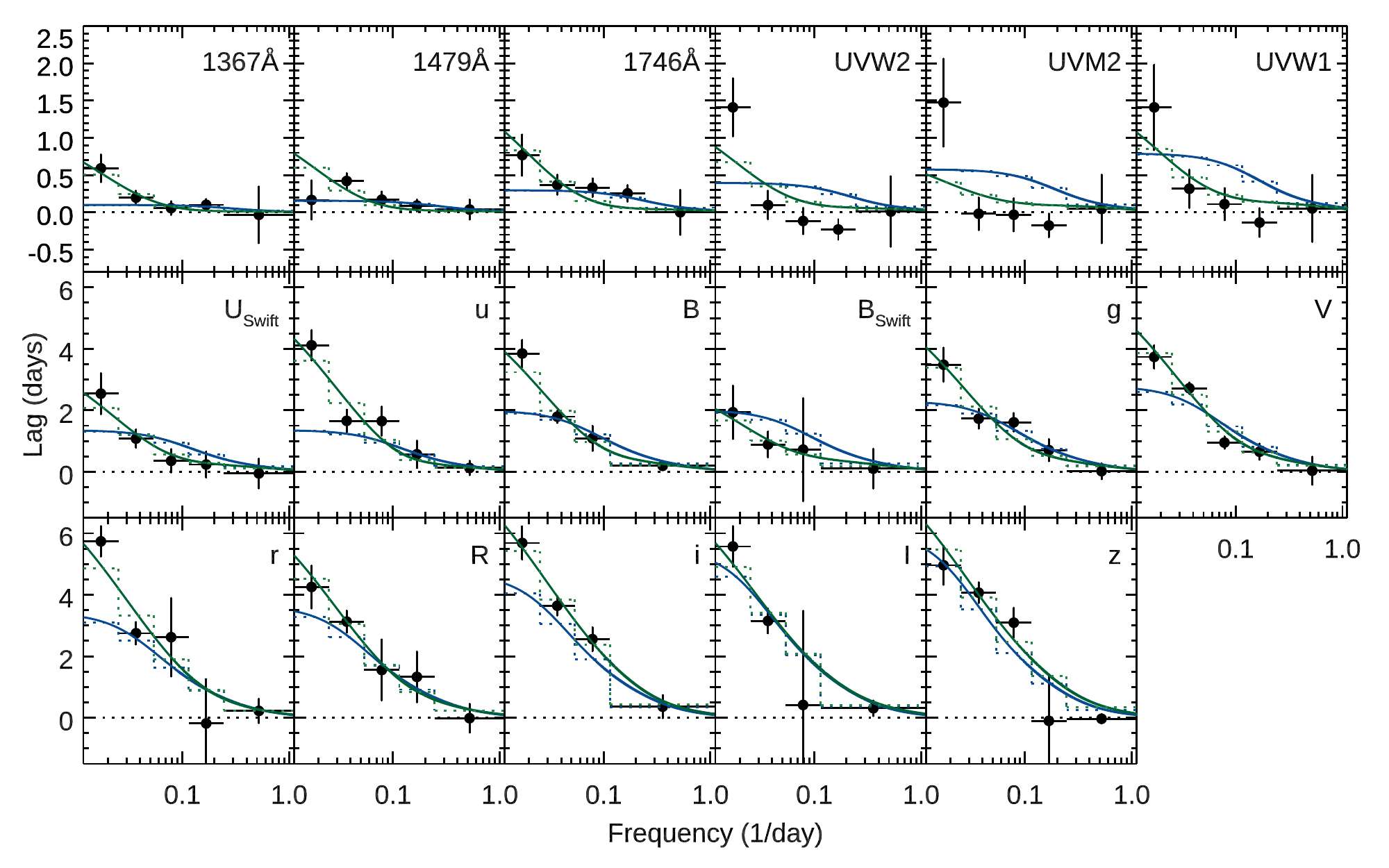}
\caption{Frequency-resolved lags in NGC 5548 with different reprocessing models.  The blue lines show reprocessing from an accretion disk only while the green lines show a model consisting of a combination of significantly smaller disk and an extended reprocessor (dotted lines show the model averaged over each frequency bin), see the text for more details. }
\label{fig:lagfreqmo}
\end{figure*}

The resulting lags from this disk model are shown as blue lines in Fig.~\ref{fig:lagfreqmo}.  While the model does well at matching the lag-frequency spectrum in the $I$ and $z$ bands, at shorter wavelengths the disk model consistently under predicts the lag at the lowest frequency and  at the shorter wavelengths over predicts the lag at the highest frequencies. In all but the longest wavelength bands, the disk model starts to flatten and only decreases slightly in lag below frequencies of around 0.1 day$^{-1}$.

\begin{figure}
\centering
\includegraphics[width=0.99\columnwidth]{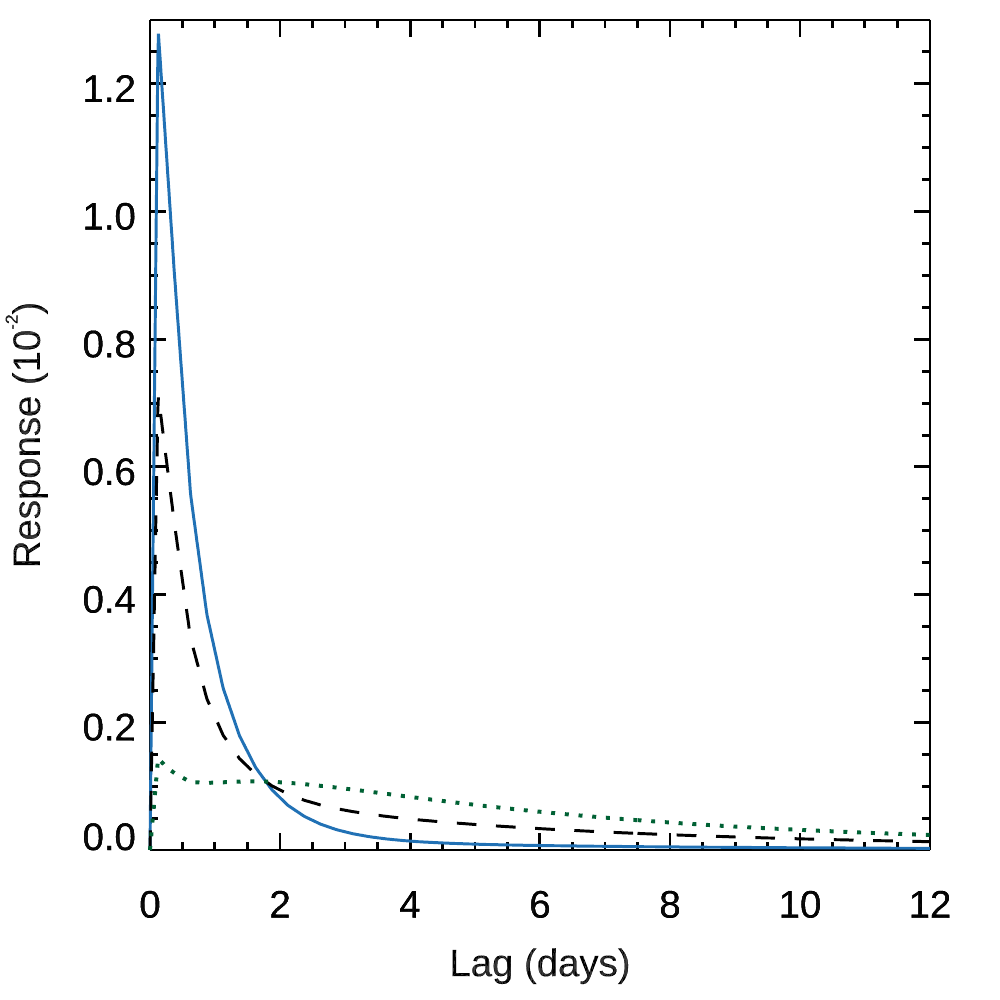}
\caption{Examples of the combined disk and BLR model response function for different values of the BLR fraction, $f$.  We use the disk response for the $V$ band combined with the BLR model with $f = 0.1$ (blue solid line), $f = 0.5$ (black dashed line), $f=0.9$ (green dotted line).}
\label{fig:tf_example}
\end{figure}

In order to get longer lags at low frequencies and shorter lags at higher frequencies we test a model for two reprocessors with reverberation from a significantly smaller disk and reverberation from a more distant, BLR-like reprocessor.  The best-fitting lag-wavelength relation from \citet{fausnaugh16} had a normalization about a factor of 3 larger than expected for reasonable assumptions about mass and accretion rate.  Our first disk model test shows that the lag-frequency spectrum is not consistent with this.  For this second model we therefore adopt a disk model that is the size expected for a standard disk.  \citet{fausnaugh16} state such a disk model would have a normalization leading to a lag at the reference band of 0.14 days.  We adjust $T_B$ and $T_F$ to match this, finding $T_B = 6400$ and $T_F = 5000$~K. Again, we assume an inclination of $45^\circ.$  In addition to this disk model, we add a log-normal shaped response (Eq.~\ref{eq:lgnorm}) to account for continuum emission from the BLR.  During the AGN STORM campaign, the H$\beta$ lag was measured to be approximately 6 days when using the 1158\AA\ reference band \citep{pei17}.  Since the median of a log-normal distribution is $e^M$, we therefore assume $M = \ln{6}$, and to give reprocessing over an extended region assume $S=1.0$.  We note these are just adopted for demonstrative purposes.  These two responses are then added together as follows:
\begin{equation}
\psi_2(t) = (1-f)\psi_{\rm disk}(t) + f\psi_{\rm BLR}(t)
\end{equation}
where $\psi_{\rm disk}(t)$ is the disk response and $\psi_{\rm BLR}(t)$ the BLR response which are both normalized to have a total area of 1.  $f$ is then the response fraction from the BLR and $\psi_2(t)$ is combined response.  We show three examples of this combined response function in Fig.~\ref{fig:tf_example}.  When $f = 0.1$ (blue solid line) the response is dominated by the accretion disk component, and has a prominent peak very close to zero.  At the other extreme, when $f=0.9$ (green dotted line) the response is dominated by the extended component, and has a more prominent tail at long lags.

\begin{figure}
\centering
\includegraphics[width=0.99\columnwidth]{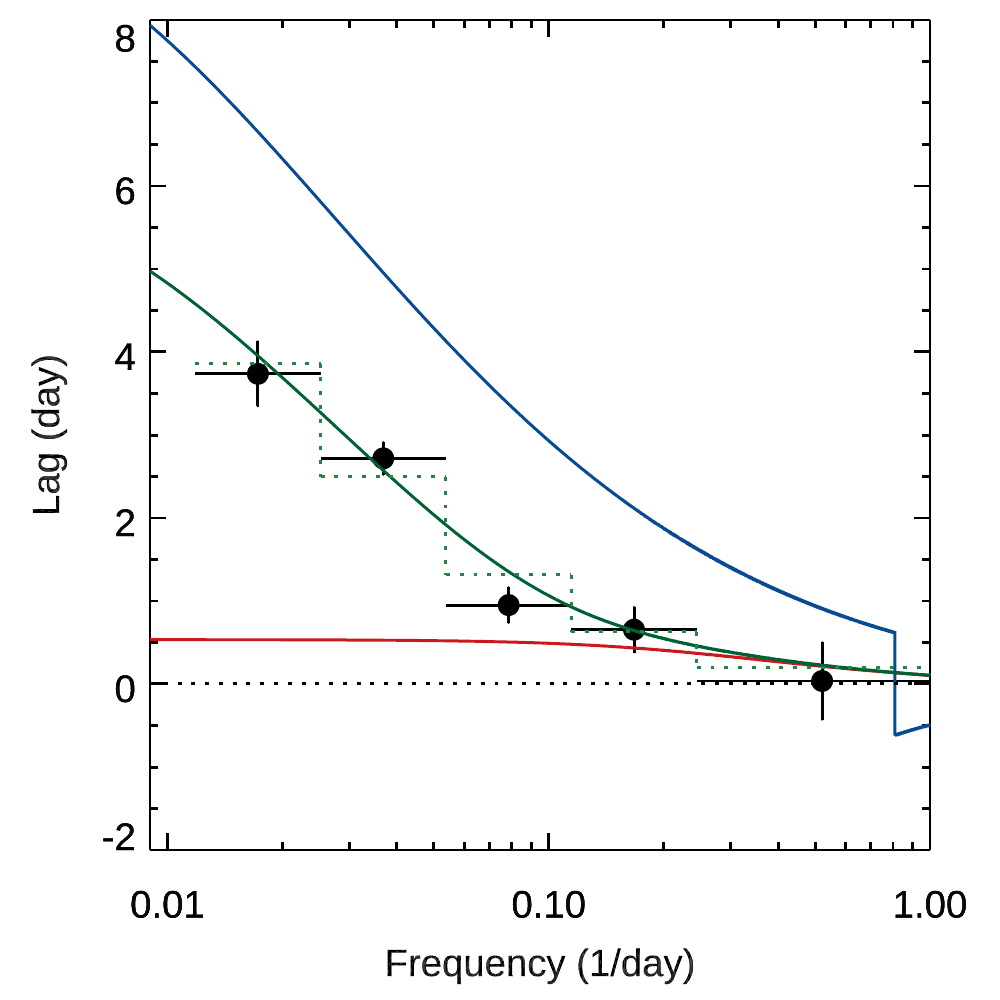}
\caption{Frequency-resolved lags in the $V$ band.  The green line shows the best-fitting disk+BLR model.  The individual components are shown as red (disk) and blue (BLR) lines.}
\label{fig:vmodel}
\end{figure}

We apply this model to the data, fitting for $f$ in each band.  The resulting lag-frequency models are shown as green lines in Fig.~\ref{fig:lagfreqmo}. This second model does a much better job of matching the observed frequency-resolved lags.  Quantitatively, the $\chi^2$ for the large disk model = 236.4 for 81 data points, giving $\chi_\nu^2 = 2.92$.  For the small disk plus BLR model we get $\chi^2 = 70.95$ for 81 data points and 17 free parameters ($f$ in each band), giving $\chi_\nu^2 = 1.11$, i.e. a significantly improved fit compared to the large disk model.  The second model does better since the short disk lags give short lags above frequency of 0.1 day$^{-1}$, while the BLR response allows the lags to match at the lowest frequencies and gives a smooth decrease in lag with increasing frequency.  A breakdown of the model components is shown for the $V$ band in Fig.~\ref{fig:vmodel}.  The disk model (red line) provides a short ($\sim0.25$ day) lag at all observed frequencies, while the BLR model (blue line)  shows decreasing lag with increasing frequency up to where phase-wrapping takes place.  The combination of the two with $f=0.63$ matches the observed lags.  

\begin{figure}
\centering
\includegraphics[width=0.99\columnwidth]{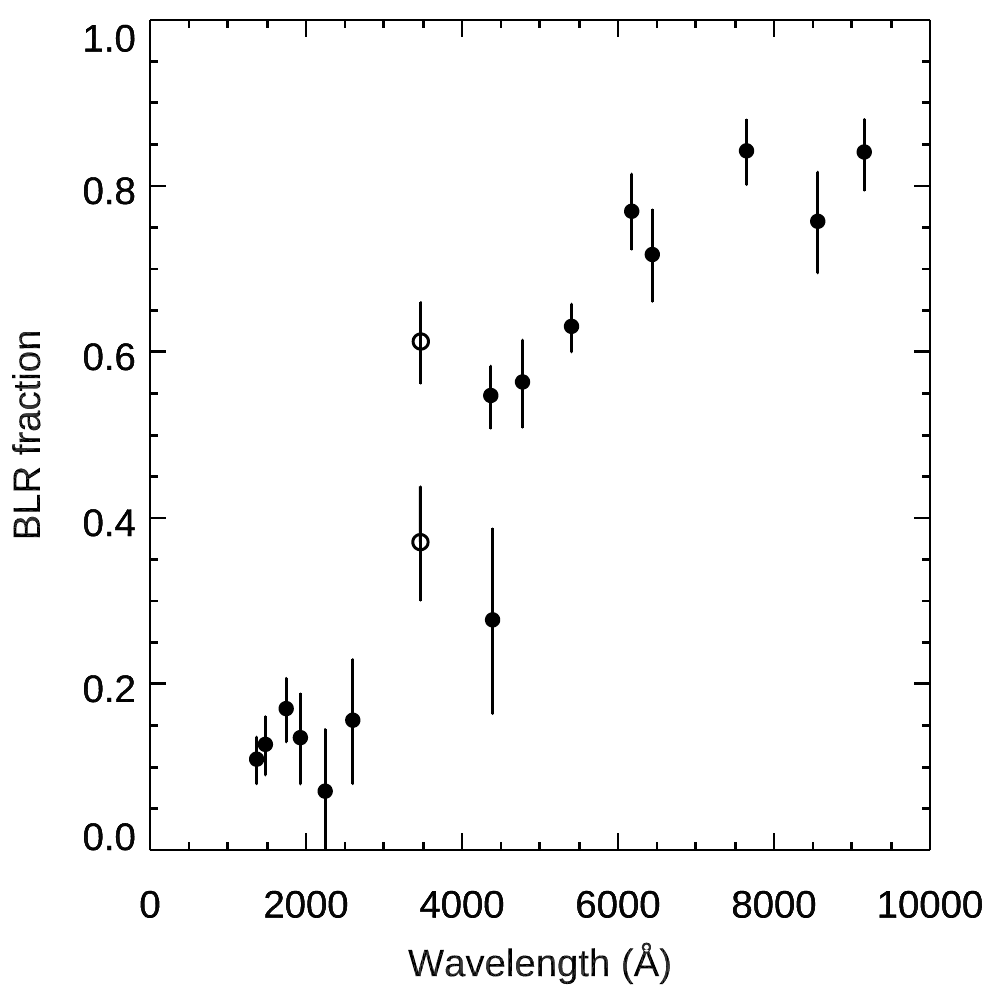}
\caption{Fraction of the response from the BLR component from the two-component (small disk plus BLR) model.  The Swift $U$ and ground-based $u$ bands are shown as open circles.}
\label{fig:blrf}
\end{figure}

Since the disk lags are short ($<1.5$ day in all bands), the BLR fraction has to increase significantly with increasing wavelength, which we show in Fig.~\ref{fig:blrf}.  This shows a generally increasing fraction with increasing wavelength. The $u$ band offset from the general trend, as would be expected if this distant reprocessor component is associated with diffuse continuum from the BLR \citep{koristagoad01}, though we note the Swift $U$ does not show a significant offset.  The $u/U$ bands are shown as open circles in Fig.~\ref{fig:laglambda} and \ref{fig:blrf}.  

For this two-component model we also tried allowing $M$ and $S$ to be free parameters in the fit, but common for all wavebands.  We explore the range for $e^M$ from 0.5 -- 11.5 days. and $S$ from 0.1 to 5.5.  We find a best-fit with parameters (and 1$\sigma$ uncertainties): $e^M = 7.3_{-0.9}^{+0.7}$ days and $S = 1.1 \pm 0.2$, this gives $\chi^2 = 68.5$ for 62 degrees of freedom.  There are strong constraints on the lower limit for $e^M$, with a 3$\sigma$ lower confidence limit of 5.0 days.  The upper-range is not as tightly constrained, with the best-fit at $e^M = 11.5$ days at approximately the 2$\sigma$ confidence level.  Again, this indicates that this more distant component is consistent with the size-scale of the BLR in NGC~5548.

We also tested fitting the lag-frequency spectra with a single log-normal component, allowing $M$ and $S$ as free parameters in each waveband.  While this does give good fits at each waveband we find that in the UV bands $M$ tends to be as small as allowed and $S$ wants to be large (in the range 4 -- 5 for the shortest wavelength bands).  This gives a response function that peaks as close to zero as possible yet has a long tail.  This is needed to have the lags close to zero at most frequencies but still give a non-zero lag at the lowest frequencies.  This is equivalent to our two-component model described above with a small disk and larger reprocessor.

To summarize, the modeling we have performed demonstrates that the frequency-resolved continuum lags in NGC 5548 are more consistent with a standard-sized accretion disk with an additional contribution from BLR continuum emission than emission from a larger than expected accretion disk alone.

\section{Discussion}

We have introduced the use of frequency-resolved lag analysis in UV/optical continuum reverberation mapping.  Since this method determines the lag at each Fourier frequency, it separates out the timescale (and hence the corresponding size scale) at which the light curves are lagging each other.  While this information is also encoded in the auto and cross correlation functions, it is not easy to assess with standard cross correlation analysis.  

To demonstrate the usefulness of frequency-resolved lag analysis we simulated light curves assuming two different reprocessing geometries - one that has a response function that is close to symmetrical, and another that has a peak at shorter lags but a significant tail to longer lags.  Standard cross correlation lag analysis recovers centroid lags for the two different light curves that are consistent with each other. However, when using frequency-resolved lag analysis they can be easily distinguished.

We apply frequency-resolved lag analysis to the UV/optical continuum light curves of NGC 5548 from the AGN STORM campaign \citep{edelson15,fausnaugh16}.  This campaign spanned approximately 170 days with daily monitoring by HST and an average of 1.5 observations per day with Swift, as well as ground-based monitoring leading to light curves spanning from 1158\AA\ to $z$.  Since the light curves are not strictly evenly sampled and do contain occasional gaps we applied the maximum likelihood method of \citet{zoghbi13}, resulting in lag measurements in 4 or 5 frequency bins (depending on the quality of the light curve and sampling) over the frequency range 0.012 to 1.11 ${\rm day}^{-1}$.  The lags are consistently longest in the lowest frequency bin and through the optical they decrease with increasing frequency.  In the UV many of the lags outside the lowest frequency are consistent with zero.  This is similar to the findings of \citet{griffiths21} who look at the lag between X-rays and UVW2 in Mrk 335, only finding tentative evidence for non-zero lags in the lowest frequency bins.

The lags do not evolve dramatically over the frequency range 0.1 -- 1 day$^{-1}$, which indicates that reprocessing on size scales less than 10 light days is occurring promptly.  The lag signal detected by CCF analysis corresponds to continuum emission happening on size scales greater than this. The lags in the frequency bin centered at 0.037 ${\rm day}^{-1}$ approximately follow a $\tau \propto \lambda^{4/3}$ relation that is consistent with the lags recovered from cross correlation analysis by \citep{fausnaugh16}.  Thus, the cross correlation analysis is picking out the lags occurring on timescales of around 27 days.  

We use accretion disk response functions of \citet{cackett07} calculated to match the observed $\tau \propto \lambda^{4/3}$ relation from cross correlation analysis.  The frequency-resolved lags from these response functions do not match the observed frequency-resolved lags well through the UV and much of the optical, though do better at the longest wavelengths.  They show lags that flatten off at the lowest frequencies and generally fail to match the lag in the lowest frequency bin.  This suggests the need for an additional reprocessor at large distances.  

As an alternative model we try a significantly smaller disk (consistent with expections for a \citealt{shakurasunyaev} disk), with an additional extended reprocessor that is a log-normal distribution with median lag of 6 days, consistent with the location of the BLR during the AGN STORM campaign \citep{pei17}.  We add the responses together but allow the fraction from the BLR component to vary from band to band.  We find that this provides a significantly better fit to the frequency-resolved lags, showing the smoothly decreasing lag with increasing frequency.  This model shows an increasing BLR fraction with increasing wavelength.

We note that it is possible to use a single log-normal response function to fit the lags.  However, in order to fit the UV lag-frequency data there needs to be a prompt response plus an extended tail to give zero lag at most frequencies yet still have a non-zero lag in the lowest frequency bin.  This is not needed for the longer wavelength bands, leading to vastly different values for $M$ in each band, which is not easy to interpret physically. On the other hand, the advantage of the two-component disk plus BLR model is that the BLR has the same response in each waveband, and the difference in lag measured depends on what fraction of the variable flux comes from the disk (prompt response) versus the BLR (extended response).

We stress that we have not explored the range of possible models in full here, and have only tested a couple of very specific scenarios.  The fits presented here do not allow the disk properties to be free parameters, moreover, we use a very simple analytic prescription to represent the BLR.  Future work involving a fuller exploration of disk models, for instance higher inclination disks have response functions that peak more prominently at short lags \citep[e.g., see fig.~3 from][]{starkey16}.  Moreover, an investigation of more complex disk models that fully take into account general relativity \citep{kammoun19,kammoun21a,kammoun21b}, and including more realistic, physically motivated BLR models \citep[e.g.,][]{koristagoad19,netzer20} is warranted.  In particular, the lag models will also need to be consistent with the spectral energy distribution of the variable component as determined by recent flux-flux analysis \citep[e.g.,][]{mchardy18,cackett20,hernandezsantisteban20}.  Regardless of the details of the model, the frequency-resolved lags demonstrate that the reprocessing gas has both a prompt response (indicated by the short lags at frequencies above 0.1 day$^{-1}$ and the lack of phase-wrapping) and an extended response over a range of radii that gives a smooth increase in lag to lower frequencies. 

The obvious physical origin for this extended response is diffuse continuum emission from the BLR \citep{koristagoad01,koristagoad19,lawther18,netzer20,netzer21}.  As described in the Introduction, the excess lag in the $u/U$ bands and around the Balmer jump \citep[e.g.,][]{edelson19,cackett18} strongly suggests that there is a significant contribution to the variable flux from the BLR diffuse continuum (however, this excess does not show up as strongly in the frequency-resolved lags here).  Moreover, analysis that removes long-term variability via subtracting off a moving box-car average effectively acts as a high-pass filter and finds shorter lags \citep{mchardy14,mchardy18}, consistent with what we find here.  Further evidence for significant BLR diffuse continuum comes from the decoupling of the continuum variations in NGC 5548 during the AGN STORM campaign \citep{goad19}.  The emission lines in NGC~5548 were observed to show anomalous behavior where they decorrelated from the UV continuum variations for a period of 60-70 days \citep{goad16}. Flux-flux analysis of the continuum variations show that they also show anomalous behavior during the same period, indicating a significant contribution to the continuum from the BLR \citep{goad19}. \citet{netzer21} demonstrates that both the shape and normalization of the observed lag-spectra are consistent with BLR diffuse emission with a covering factor of $\sim$0.2.

Our model of a small disk and extended reprocessor whose flux fraction increases with wavelength supports the findings of \citet{chelouche19}.  They perform a bivariate lag analysis of Mrk 279 assuming the variations consist of a driving component that varies nearly simultaneous across the optical, and an additional delayed component that partly contributes to each band. They find the fraction of variable flux from the delayed component increases with wavelength and dominates at the longest wavelengths.  This is consistent with our best-fitting model where the response from the disk gives a very short lag at all wavelengths, and the fraction of response from the extended region increases with wavelength and is dominant in the reddest bands.

The observed frequency-resolved lag behavior can also be used to test alternative models for the continuum lags.  In the corona-heated accretion-disk reprocessing (CHAR) model  \citep{sun20a,sun20b} magnetic fluctuations in the corona affect the disk MHD turbulence and heating rate driving temperature changes in the disk.  The frequency-resolved lags from that model (see fig. 7 of \citealt{sun20a}) decrease smoothly with increasing frequency, broadly consistent with what we observe in NGC~5548.

Over the last 5 years or so a sample of 7 additional AGN have published intensive accretion disk reverberation mapping campaigns involving Swift with similarly high cadence datasets to NGC 5548 \citep{edelson17,edelson19,cackett18,cackett20,mchardy18,hernandezsantisteban20,vincentelli21,kara21}, and several more are either on-going or recently completed.  In future work we will explore frequency-resolved lags in those datasets to see whether they also display the same shaped lag-frequency spectra.  Furthermore, this technique can be applied to emission-line reverberation too as an additional way to assess the shape of the response function. In this case, it will be important to take account of the response function between the driving light curve and the 5100\AA\ emission to properly assess the frequency-resolved lags.  For continuum reverberation mapping the fact that the frequency-resolved lags are longest at the lowest frequencies, drives the need to want as long a monitoring campaign as possible. Only when the frequency-resolved lags flatten off have we probed the longest timescales where the gas is no longer responding.  Similarly, for emission-line reverberation longer campaigns will better capture the full response of the BLR.  This should be considered when planning future campaigns.

\begin{acknowledgements}
We thank Keith Horne for comments that helped improve the paper. EMC gratefully acknowledges support from the NSF through grant AST-1909199. OU is grateful for support from the NSF through PHY-1460853 that funded the Research Experience for Undergraduates program at Wayne State University where this project initially began. 
\end{acknowledgements}

\bibliographystyle{aasjournal}
\bibliography{agn}

\end{document}